\documentclass[11pt, hyphens]{kimreview}

\usepackage{bm}
\usepackage{amsmath}
\usepackage{amssymb}
\usepackage{epstopdf} 
\usepackage{wrapfig}  
\usepackage{url}
\usepackage{xpatch}
\usepackage{dcolumn}
\newcolumntype{d}[1]{D{.}{.}{#1}} 

\usepackage[style=numeric-comp,sorting=none,backend=biber,maxbibnames=50,giveninits=true]{biblatex}

\usepackage{hyperref}
\usepackage{xurl}
\hypersetup{breaklinks=true}
\renewbibmacro{in:}{}

\DeclareBibliographyAlias{article}{custom}
\DeclareFieldFormat{pages}{#1}
\DeclareFieldFormat{url}{\url{#1}}
\DeclareBibliographyDriver{article}{%
  \usebibmacro{bibindex}%
  \usebibmacro{begentry}%
  \usebibmacro{author/editor+others/translator+others}%
  \setunit{\labelnamepunct}\newblock
  \usebibmacro{title}%
  \newunit\newblock
  \usebibmacro{byauthor}%
  \newunit\newblock
  \usebibmacro{byeditor+others}%
  \newunit\newblock
  \printfield{version}%
  \newunit\newblock
  \printfield{journaltitle}%
  \newunit\newblock
  \iffieldundef{volume}
    {}
    {\printfield{volume}}
  \iffieldundef{number}
    {}
    {\printtext{(\printfield{number})}}
  \iffieldundef{pages}
    {}
    {\printtext{:\printfield{pages}},}
  \newunit\newblock
  \printfield{year}
  \newunit\newblock
  \iftoggle{bbx:eprint}
    {\usebibmacro{eprint}}
    {}%
  {\printtext{\printfield{url}}}
  \usebibmacro{finentry}%
}

\DeclareBibliographyDriver{misc}{%
  \usebibmacro{bibindex}%
  \usebibmacro{begentry}%
  \usebibmacro{author/editor+others/translator+others}%
  \setunit{\labelnamepunct}\newblock
  \usebibmacro{title}%
  \newunit\newblock
  \usebibmacro{byauthor}%
  \newunit\newblock
  \printfield{version}%
  \newunit\newblock
  \newunit\newblock
  \printfield{journaltitle}%
  \newunit\newblock
  \iffieldundef{volume}
    {}
    {\printfield{volume}}
  \iffieldundef{number}
    {}
    {\printtext{(\printfield{number})}}
  \newunit\newblock
  \iffieldundef{pages}
    {}
    {\printtext{:\printfield{pages},}}
  \newunit\newblock
  \newunit\newblock
  \iftoggle{bbx:eprint}
    {\usebibmacro{eprint}}
    {}%
  \newunit\newblock
  \usebibmacro{url+urldate}%
  \newunit\newblock%
  \usebibmacro{addendum+pubstate}%
  \setunit{\bibpagerefpunct}\newblock
  \usebibmacro{pageref}%
  \newunit\newblock
  \iftoggle{bbx:related}
    {\usebibmacro{related:init}%
     \usebibmacro{related}}
    {}%
  \usebibmacro{finentry}%
}

\DeclareFieldFormat{labelnumberwidth}{#1}
\DeclareFieldFormat{shorthandwidth}{#1}

\DeclareRangeChars{-}
\DeclareRangeCommands{\bibrangedash}

\DeclareCiteCommand{\cite}[\mkbibbrackets]
  {\usebibmacro{cite:init}%
   \usebibmacro{prenote}}
  {\usebibmacro{citeindex}%
   \usebibmacro{cite:comp}}
  {}
  {\usebibmacro{cite:dump}%
   \usebibmacro{postnote}}

\begin{document}


\title{From GAP to ACE to MACE}


\authorone{Noam Bernstein}
\affiliationone{Center for Materials Physics and Technology, U.~S. Naval Research Laboratory, Washington, D.C.}



\publishyear{2024}
\volumenumber{2}
\articlenumber{05}
\submitdate{October 1, 2024}
\publishdate{October 7, 2024}
\doiindex{10.25950/67c762ea}
\doilink{10.25950/67c762ea}

\paperReviewed
{Gaussian Approximation Potentials: The Accuracy of Quantum Mechanics, without the Electrons}
{A.~P. Bart\'ok, M.~C. Payne, R. Kondor, and G. Cs\'anyi}
{\href{https://doi.org/10.1103/PhysRevLett.104.136403}{Phys.\ Rev.\ Lett., 104:136403 (2010)}}

\paperReviewedTwo
{MACE: Higher Order Equivariant Message Passing Neural Networks for Fast and Accurate Force Fields}
{I. Batatia, D.~P. Kov\'acs, G. Simm, C. Ortner, and G. Cs\'anyi}
{\href{https://proceedings.neurips.cc/paper_files/paper/2022/hash/4a36c3c51af11ed9f34615b81edb5bbc-Abstract-Conference.html}{Adv. Neural. Inf. Process. Syst., 35:11423 (2022)}}


\maketitle



\begin{abstract}

The Gaussian approximation potential (GAP) machine-learning-inspired
functional form was the first to be used for a general-purpose
interatomic potential.  The atomic cluster
expansion~\cite{drautz_atomic_2019,drautz_erratum_2019} (ACE),
previously the subject of a KIM Review~\cite{ortner_atomic_2023},
and its multilayer neural-network extension (MACE) have joined GAP among
the methods widely used for machine-learning interatomic potentials. Here
I review extensions to the original GAP formalism, as well as ACE and MACE-based
frameworks that maintain the good features and mitigate the limitations of the
original GAP approach.

\end{abstract}
\medskip

\section*{Background: Gaussian Approximation Potentials with Smooth Overlap
         of Atomic Positions (SOAP-GAP)}

Machine learning interatomic potentials (MLIPs) have revolutionized
the field of atomistic simulations by replacing functional forms
that are motivated by the specific physics of the bonding between
atoms with minimally constrained many-body 
forms, as reviewed in \cite{bartok_machine_2017,deringer_machine_2019,behler_machine_2021,deringer_gaussian_2021,musil_physics-inspired_2021}.
Since any sufficiently flexible functional form could (by definition)
reproduce an arbitrarily complicated potential energy surface (PES),
one of the main practical challenges in developing MLIPs has been the
tradeoff between this flexibility and the resulting tendency
for inaccuracy away from the fitting data.  If we had reference
data available for all configuration (an infinite amount), that
would fully constrain any fitting function.  However, since generating
reference data (e.g.\ with density functional theory calculations)
is computationally expensive, reducing the amount of required data
is essential.
This can be achieved by taking advantage of features of the PES,
such as symmetries or smoothness, that reduce the possible variation
of the energy between fitting configurations, and hence the amount
of data needed to constrain the fit. This is particularly important
for the very flexible functions required to produce accurate and
general purpose MLIPs that are applicable throughout a wide range
of configurations and geometries.

About 15 years after the earliest proposals to use a machine-learning method,
neural networks, for
interatomic potentials~\cite{skinner_neural_1995}, the first modern
MLIPs succeeded in describing 
atomic geometries for specific physical systems~\cite{behler_generalized_2007,bartok_gaussian_2010}.
The Gaussian approximation potential (GAP) approach~\cite{bartok_gaussian_2010}
uses sparse Gaussian process regression with a number of different descriptors
of the atomic environments.
One of the first efforts to develop a {\em general purpose} MLIP
for a particular material was the silicon GAP~\cite{bartok_machine_2018}
that used smooth overlap of atomic positions (SOAP)
descriptors. This potential was
able to reproduce the energies of its density functional theory
(DFT) reference energies (and their force and stress gradients) for
a single element in a wide range of crystal structures, defects
(point, line, and plane), and disordered states such as liquid and
amorphous.  The SOAP-GAP functional form had exact rotation and
permutation symmetry, as well as smoothness, due to the form of the
squared exponential SOAP descriptors, and regularization (reducing
the tendency to follow every fluctuation in the fitting data) through
the ``noise'' hyperparameters of Gaussian process
regression~\footnote{Note
that most of this so-called noise is in fact model error due to the
finite range of the potential. Actual uncertainty in the reference
values, e.g.\ due to incomplete convergence, is much smaller.}.
The flexibility of the function could be controlled through the SOAP
descriptor parameters and the number of representative points in
the sparse GPR.

This pioneering effort showed that an accurate general purpose MLIP
could be developed, but the specific methods that made it possible
also led to a number of limitations.  The computational cost was
large for an interatomic potential, of order 10-100~ms/atom on a
single CPU core~\cite{zuo_performance_2020}, and it scaled with the
number of chemical elements squared.  
The use of GPR made the actual model fitting trivial by reducing
it to the computation of the pseudo-inverse of the covariance matrix
(via numerical linear algebra), but there was only mixed success
in the initial attempts to use the GPR-predicted variance as a
predictor of the MLIP error.
%
Here I briefly survey some of the improvements and successors to SOAP-GAP
that attempted to maintain the advantages and mitigate the disadvantages,
while leveraging other mathematical and computational developments.

\section*{Improvements to SOAP-GAP}

One change that minimally altered the overall SOAP-GAP form but led
to substantial speedup was the development of new
descriptors~\cite{caro_optimizing_2019} implemented in the {\tt
soap\_turbo} package~\cite{noauthor_libatomssoap_turbo_2022}.  The SOAP kernels
represent the similarity of two atomic environments (i.e.\ the kernel
for GPR) starting from a spherically symmetric Gaussian centered
around each neighboring atom.  In {\tt soap\_turbo}, these Gaussians
are replaced with a product of a radial Gaussian and an angular
Gaussian.  This modification gives some additional flexibility, for
example varying the smoothing, i.e. the width of the Gaussians,
with neighbor distance, and crucially it is considerably faster to
evaluate. With {\tt soap\_turbo} it is possible to match the accuracy of a
conventional SOAP-GAP at $\sim 10 \times$ lower computational
cost~\cite{caro_optimizing_2019}. As a result, these improved descriptors
have essentially replaced the original SOAP variant in the QUIP
{\tt gap\_fit} software~\cite{csanyi_libatomsquip_2021}.

A larger change, to address the quadratic scaling with the number
of chemical elements, was proposed by Darby {\em et al.} through
an approach called tensor-reduced density
representations~\cite{darby_tensor-reduced_2023}.  The SOAP-GAP expressions
were re-cast in terms of quantities defined in the atomic cluster
expansion (ACE) formalism~\cite{drautz_atomic_2019,drautz_erratum_2019}.  This redefinition
makes it clear that the expansion of the SOAP descriptors into
chemical elements and radial basis functions, whose number of tensor indices
grows linearly with the number of chemical elements, can be transformed
into an embedding~\footnote{Here ``embedding'' is used in its
machine-learning meaning, which is unrelated to the {\em emedded}
atom method (EAM) potentials.}, i.e.\ a linear transformation into a space of 
arbitrary dimensionality, which can be converged to recover the exact tensor.
The great advantage of this transformation is that the size of this
space does not grow with combinatorial complexity as in earlier formulations.
Darby {\em et al.} conclude that, in practice, the cost of the
computation for evaluating the GAP becomes independent of the
composition.
They find that the tensor-reduced representation leads to a reduction by a factor of $\sim 10$
in the number of descriptor vector elements required for a given
accuracy~\cite{darby_tensor-reduced_2023,shenoy_collinear-spin_2024}.
Further exploration
will be needed to determine the optimal size of the embedded space,
the best method to optimize the linear transformation, and under
what circumstances (e.g.\ number of elements) operating in the
embedded space leads to a net reduction in computational cost.

GPR, which underlies GAP, predicts not only the value of the fit
quantity, i.e.\ the potential energy, but also its variance, which
is a measure of its uncertainty. It was initially hoped that this
variance would be a good predictor of the MLIP's error, and this
was indeed demonstrated for the Si SOAP-GAP~\cite{bartok_machine_2018},
as well as for a SOAP-GAP fit to energy barriers in heterogeneous
catalysis~\cite{schaaf_accurate_2023}.  However, in some other systems
where SOAP-GAP was applied this was not the case, and we suspect
that the successful examples depended on the availability of a large amount
of fitting data relative to the diversity of configurations. Since
the initial demonstration, a number of efforts have been made to
improve GAP error prediction based on GPR-predicted variance.  One
approach applied to SOAP-GAP is to optimize the model's
hyperparameters~\footnote{Hyperparameters are parameters that control
the behavior of a model but are set by the developer, either {\it a priori}
or through some heuristics, in contrast to model parameters that are adjusted by
the regression method to minimize the error on the fitting data.}
by maximizing
the Bayesian log-likelihood~\cite{bartok_improved_2022}.  This is
a computationally expensive process, but this cost is partially
mitigated by a new parallelized implementation of the GAP fitting
code~\cite{klawohn_massively_2023}.
Another approach that proved successful was to speed up hyperparameter
optimization and avoid the need for sparsification of the representative
points by limiting the amount of reference data and restricting the
model to a specific system~\cite{jinnouchi_--fly_2019}, or by
limiting the model's non-linearity (reducing its body order and
potentially its accuracy)~\cite{vandermause_--fly_2020}.

\section*{Beyond GAP}

While the extensions to GAP discussed above have addressed some of
its limitations, other approaches, with their own complementary
advantages and disadvantages, are also under development.  The two
that I describe below, linear ACE as implemented in {\tt
ACEpotentials.jl} and nonlinear neural-network multilayer-ACE (MACE), were
influenced by the same ideas and motivations as the
design choices of SOAP-GAP.

\subsection*{Linear atomic cluster expansion via {\tt ACEpotentials.jl}}

As the work of Drautz~\cite{drautz_atomic_2019,drautz_erratum_2019,ortner_atomic_2023} showed,
a very wide range of interatomic potentials can be seen as aspects
of a unified atomic cluster expansion (ACE) framework.
These include SOAP-GAP, a non-linear model with 3-body
descriptors, the spectral neighbor analysis potential
(SNAP)\cite{thompson_spectral_2015,wood_extending_2018}, a linear or quadratic
model with 4-body descriptors, and a number of linear models with
arbitrary body order more directly tied to the ACE functional form.  Several
codes implementing the latter approach are available, including
moment tensor potentials (MTP)~\cite{shapeev_moment_2016}, performant
ACE~\cite{lysogorskiy_performant_2021} and
PACEMAKER~\cite{bochkarev_efficient_2022,noauthor_pacemaker_nodate}, and
the implementation I discuss here.
Combining the lessons of GAP with ACE led to the development of a new ACE
fitting implementation, {\tt ACEpotentials.jl}~\cite{witt_acepotentialsjl_2023,noauthor_acesuitacepotentialsjl_2024}.
As in SOAP-GAP, rotation and permutation symmetry
are exact (due to the ACE framework), but smoothness and regularization
are facilitated through entirely different means.

Rather than using the squared exponentials that smooth the atomic position
variation in SOAP,\linebreak {\tt ACEpotentials.jl} uses a linear model with
polynomial basis, whose oscillatory tendencies are quantified by
the polynomial degree.  A mathematical expression combining the
radial and angular oscillations defines a {\em total} polynomial
degree, whose maximum value is used to truncate the basis, with the
possibility of separate control over radial and tangential variation.
The fitting is done using Tikhonov or ridge regression, which adds
a term to the minimized loss function related to the magnitude of
the coefficients of each basis polynomial.  This term is intended
to make the function smooth by more strongly penalizing large coefficients,
since they are likely to lead to unphysical oscillations away from
the fitting data.  Particular
choices for the weight of each basis function's coefficient in the
ridge term can maximize particular quantifications of smoothness,
such as the average curvature, or reproduce the smoothness of the
SOAP squared-exponential kernels.  

Ridge regression can also be analyzed through a Bayesian interpretation,
where the ridge term can be thought of as a Bayesian prior favoring
smooth fits.  This formulation makes it possible to automatically
optimize the level of regularization on a global (Bayesian ridge
regression, BRR) or per-coefficient (automatic relevance determination,
ARD) basis~\cite{witt_acepotentialsjl_2023}, and to produce an
ensemble, or committee, of potentials for uncertainty quantification by drawing
from the probability distribution of polynomial
coefficients~\cite{van_der_oord_hyperactive_2023}.

Like GAP, the scaling of {\tt ACEpotentials.jl} potentials with the number
of elements is a power law (with an exponent that depends on body
order). Through the same analysis as for GAP, Darby {\em et al.}
applied tensor reduction to ACE and showed similarly promising
results, but these have not yet been explored in a wide range of
applications, or incorporated into the current generally available
implementation.

As its name suggests, the fitting framework of {\tt ACEpotentials.jl} is
implemented in Julia~\cite{bezanson_julia_2017}, and is mainly used in that
environment.  To use the resulting potentials, an atomic-simulation
environment (ASE) calculator is available, and most, although not
all, of the features can be exported into the format used by the
performant ACE (PACE) implementation~\cite{lysogorskiy_performant_2021}.  This high speed
evaluation code can be used within the LAMMPS simulation
software~\cite{thompson_lammps_2022} for a wide range of time propagation and sampling
methods with parallelism based on spatial domain decomposition.

\subsection*{Nonlinear neural-network multilayer atomic cluster expansion (MACE)}

The ACE formulation was used as the basis for a number of multi-layer
neural-network architectures, including NequIP~\cite{batzner_e3-equivariant_2022} and
multi-ACE~\cite{bochkarev_multilayer_2022}. Here I discuss a particular special
case of the more general multi-ACE approach, MACE, which is a very flexible non-linear extension
of the ACE framework~\cite{batatia_mace_2022,noauthor_acesuitmace_2024}.  Its
input is the ACE description of the atomic environment, starting
from an atomic basis, forming a tensor product basis, and finally
a rotationally invariant symmetrized basis~\cite{kovacs_evaluation_2023}.  Rather than
using it as a basis for a linear model as in {\tt ACEpotentials.jl}, this
description is fed into an equivariant message-passing graph neural
network (GNN) that couples messages, constructed from symmetrized-basis
tensors, along edges between nodes corresponding to neighboring
atoms.  Each layer constructs output node features based on input
messages from neighboring atom nodes, and the final, readout, layer
predicts output energies (or, in principle, other tensor quantities)
based on the final node features.  As with all neural networks,
there is the potential for extreme flexibility due to the range of
options for each of these stages.

In practice, some simple choices for the details of the architecture
have led to remarkably accurate and stable MLIPs across a vast range
of materials.  In general the tensor product basis is truncated to
body order 4, two GNN layers are used, and the messages include
angular momentum components of at most 2. These choices limit the
nonlinearity to the tensor product basis construction, and to the
radial basis functions and energy readout function, which are
implemented as fully connected NNs (multilayer perceptrons, MLPs).
This mild degree of nonlinearity is meant to strike an optimal balance
between functional form flexibility on the one hand, and ease
of fitting on the other.

The most remarkable result is that, in addition to making more
accurate material-specific MLIPs than GAP and {\tt
ACEpotentials.jl}~\cite{kovacs_evaluation_2023,schaaf_accurate_2023,baldwin_dynamic_2024}, MACE
has demonstrated the ability to create a universal foundation model
that is applicable across essentially the entire periodic table.
MACE-MP0~\cite{batatia_foundation_2024}, based on the Materials Project (MP)
database~\cite{jain_commentary_2013,noauthor_materials_nodate},
and MACE-OFF23~\cite{kovacs_mace-off23_2023}, based
mainly on the SPICE database~\cite{eastman_spice_2023}, have demonstrated
accurate results and stable dynamics and sampling across extremely
wide ranges of configuration space.  MACE-MP0 achieves good accuracy
across 89 elements in structures represented in the MP project
(about 20~meV/atom energy MAE and 45~meV/\AA\ force MAE for the
medium model), and at least qualitatively reasonable description
for 30 tested applications far beyond the configurations present in the fitting data.
While the precise properties it predicts
are often approximate, it is nearly always stable enough to carry
out dynamics that can be used to efficiently generate new fitting
configurations for fine-tuning. MACE-OFF23, fit to organic molecules
containing 10 elements, is even more accurate (about 1-2~meV/atom
energy RMSE and 20-30~meV/\AA\ force RMSE), although unlike MACE-MP0
its fitting and testing data consist almost entirely of equilibrium
configurations.

The MACE potential is currently implemented~\cite{noauthor_acesuitmace_2024}
using the {\tt pytorch} library~\cite{ansel_pytorch_2024}, with a python script for fitting,
an {\tt ASE Calculator} interface for evaluation from {\tt python},
and a {\tt torchscript}-based interface for use in LAMMPS~\cite{thompson_lammps_2022}.

\section*{Direct comparison}

To give a more concrete idea of the relative accuracy and speed of
these MLIPs, I fit GAP, ACE, and MACE models to a database of
Cu$_{x}$Al$_{1-x}$ DFT calculations.  Note that the comparison below
is meant to give an overall sense of their relative performance, with
the caveat that these results are based on somewhat arbitrary heuristics
for the various hyperparameters, and could vary significantly for
other choices.

The fitting set consists of 4587 configurations from an iterative
random-structure search and GAP fitting
process~\cite{bernstein_novo_2019,bernstein_how_nodate} supplemented
with automatically generated supercells, including various defects,
at a range of temperatures up to and including the liquid phase.
An additional 682 configurations from the later iterations of this
process were reserved for testing.  Reference energies, forces, and
stresses were computed using DFT-PBE~\cite{perdew_generalized_1996} with
the VASP software~\cite{kresse_efficient_1996}.
A GAP model was fit using the default heuristics of {\tt
gap\_rss\_iter\_fit}~\cite{bernstein_how_nodate}, with two SOAP
descriptors per element, and a total of 4000 sparse points.  
Two ACE models were fit using the GAP cutoff of 6.5~\AA: a faster
potential with body order 3 and maximum polynomial degree 10 (338
basis functions), and a slower potential with body order 5 and
polynomial degree 12 (5370 basis functions).  
Finally, two MACE models were fit by
fine-tuning (with multihead stabilization) the small and
medium foundation models, which differ in hyperparameters affecting
their speed, memory use, and accuracy, from
Ref.~\cite{batatia_foundation_2024}.  These notably differ
from GAP and ACE in their effective interaction range of 12~\AA.

In Table~\ref{table:pot_comparison}, I show the error on the fitting
and testing databases and the computational cost of the potential
on a single CPU core (GAP, ACE) or GPU (MACE) in a 1024~atom molecular-dynamics (MD) simulation 
using LAMMPS~\cite{thompson_lammps_2022}. This system size is
chosen to be large enough that the GPU-related
latency is a small contribution and the listed times will be roughly
constant for larger systems.  All three MLIPs produce accurate
models, with errors that, in our experience, correspond to useful
accuracy (of order 5-10\%) for physical properties such as binding
energies, elastic constants, defect formation energies, etc.  The
faster ACE model is considerably less accurate than GAP, while the
slower one matches GAP in accuracy.  Both MACE models have very
similar accuracy, matching or exceeding that of the GAP and
ACE models.  This may be due to the nonlinearity and additional
parameters in the MACE functional form, as well as the longer cutoff.

\begin{table}
  \centering
  \caption{Root-mean-squared error for fitting and testing data
  on energy (meV/atom), forces (meV/\AA), and stress (meV/\AA$^3$),
  and computational time in ms/atom/time~step for MD of a 1024 atom
  configuration on a single Intel Xeon Gold 6226R 2.9~GHz CPU core (GAP, ACE)
  or NVIDIA A100 GPU (MACE).}
  \label{table:pot_comparison}
  \begin{tabular}{l|d{3.0}d{3.0}d{1.1}|d{2.1}d{3.0}d{1.1}|d{1.3}}
      potential   & \multicolumn{3}{c|}{fit error} & \multicolumn{3}{c|}{test error} & \multicolumn{1}{c}{time} \\
                  & \multicolumn{1}{c}{energy} & \multicolumn{1}{c}{force} & \multicolumn{1}{c|}{stress} &
                    \multicolumn{1}{c}{energy} & \multicolumn{1}{c}{force} & \multicolumn{1}{c|}{stress} & \\ \hline
      GAP         &  93    &  80   &  3.3   & 8.2    & 78    & 3.0    & 2.1 \\
      ACE 3, 10   & 140    & 140   &  6.1   & 18     & 140   & 5.3    & 0.057 \\
      ACE 5, 12   & 100    &  84   &  2.4   & 8.5    & 82    & 2.4    & 0.18 \\
      MACE small  & 95     &  86   &  1.6   & 5.8    & 76    & 1.7    & 0.042 \\
      MACE medium & 96     &  82   &  1.4   & 5.7    & 80    & 1.7    & 0.12 \\
  \end{tabular}
\end{table}

The computational speed is somewhat harder to compare, because GAP
and ACE scale nearly perfectly linearly in the number of atoms and 
their current implementations are efficiently parallelized using domain decomposition, while MACE
has a significant latency due to CPU-GPU communication and has not
yet been efficiently parallelized.  The table comparison is in
the regime of a large number of atoms ($\sim 1000$) per CPU core (GAP, ACE)
or GPU (MACE).  GAP is the slowest of the MLIPs, at about 2~ms/atom/time~step.  The
two ACE variants show some of the available tradeoff range, with
the less accurate one $\sim 35\times$ faster than GAP and the more
accurate one only $\sim 12\times$.  The two MACE potentials 
have a comparable speed on a single GPU to ACE on a single CPU core.

The constraints of the current MACE implementation to a single GPU
mean that the largest system size on a GPU with 80~GB of memory
is about 21,000 for the small model and 8,000 atoms for the medium
model.  With the small model, the largest system can be simulated
for $\sim 0.11 \times 10^6$~time-steps/day, but due to latency even
a 54 atom system can only be simulated for $\sim 8 \times
10^6$~time-steps/day, about five times slower per atom.
ACE, on the other hand, is comparable in per-atom speed on a single CPU core
to MACE on a GPU, and when run in parallel loses only about 30\% in speed
with only 100~atoms/CPU~core. Therefore, a single 128~core node 
should be able to simulate 21,000 
atoms with the faster, less accurate ACE for $\sim 7 \times
10^6$~time~steps/day, about $60 \times$ faster than MACE.  The
development of efficient CPU implementations of MACE, as well as
improvements to its domain decomposition parallel efficiency for
both the GPU and CPU versions, are ongoing.

\section*{Outlook}

Rapid advances in MLIPs are beginning to remake the field of
atomistic simulations of material properties.  The success of
SOAP-GAP showed that it was possible to develop sufficiently
expressive functional forms to describe a material's PES across a
wide range of geometries, but with sufficient smoothness that they
could be fit to a practical amount of DFT-calculated data.  New
advances in descriptors for GAP will make it possible to apply this
capability to larger and more chemically complex systems. 

ACE models can match the accuracy of GAP models, but are one
to two orders of magnitude faster, even when compared to the relatively
fast {\tt soap\_turbo} formulation.  This speedup can
enable the simulation of larger systems, for example approaching the scale of multiple
crystalline grains separated by grain boundaries (of order $10^4-10^5$
atoms), or enable longer time sampling (of order nanoseconds, $10^6$ time
steps).

MACE enables the creation of foundation models that describe nearly
all chemical elements and can be used as a basis for fine-tuned
models for a specific material system or reference data calculation
method.  This capability has already revolutionized approaches
to simulating new, chemically complex systems, including systems
as diverse as as solid state acid proton conductors, polymer-based
Li-ion sulfide electrolytes, and structural metallic alloys.

Especially by combining the best ingredients of GAP, ACE, and MACE,
as well as the many others I did not have time or space to present,
MLIPs are becoming capable of dramatically more accurate calculations
of properties for systems with chemical and structural complexity that
approaches experimentally and technologically relevant materials.

\printbibliography

\end{document}